\documentstyle[pra,aps]{revtex}
\begin{document}
\draft

\title{Inverse Mass Expansion of the One-Loop Effective Action}
\author{Alexander A. Osipov\thanks{On leave from the Laboratory 
        of Nuclear Problems, JINR, 141980 Dubna, Russia},
        Brigitte Hiller}
\address{Centro de F\'{\i}sica Te\'{o}rica, Departamento de F\'{\i}sica
         da Universidade de Coimbra, 3004-516 Coimbra, Portugal}
\date{\today}
\maketitle

\begin{abstract}
A method is described for the development of the one-loop effective 
action expansion as an asymptotic series in inverse powers of the 
fermion mass. The method is based on the Schwinger-DeWitt proper-time 
technique, which allows for loop particles with non-degenerate masses.
The case with $SU(2)\times SU(2)$ as the symmetry group is considered. 
The obtained novel series generalizes the well-known Schwinger-DeWitt
inverse mass expansion for equal masses, and is chiral invariant at
each order. We calculate the asymptotic coefficients up to the fifth 
order and clarify their relationship with the standard Seeley-DeWitt 
coefficients.  
\end{abstract}



Quite often in physics one has to extract the dominant contribution of the 
short distance effects on the large distance behavior. This is the case 
where usually effective field theories (EFT) come into play 
\cite{Manohar:1996}. Chiral perturbation theory (CHPT) in QCD is a typical 
example \cite{Weinberg:1979}. It is an effective theory approach in which 
the low-energy QCD Lagrangian is constructed as a combined derivative and 
light quark mass expansion. The expansions for the effective action 
accumulate the most important features of the short distance physics, 
i.e., for instance, the quantum fluctuations of the light particles (pions) 
in CHPT, or the low-energy effects resulting from the presence of heavy 
particles in the fundamental theory \cite{Lee:1989}. Other examples are 
given by models of the Nambu -- Jona-Lasinio (NJL) type \cite{Nambu:1961}
where the action of the respective low-energy EFT results from bosonization 
of the quark-antiquark interaction \cite{Volkov:1984}. In all the above 
cases and in many others arises the necessity of calculating the determinant 
of the positive definite elliptic operator which governs quadratic 
fluctuations of quantum fields in the presence of a definite background
and which contains in compact form all information about the one-loop 
contribution of quantum fields. This determinant may be defined using the 
Schwinger proper-time representation \cite{Schwinger:1951,DeWitt:1965} 
in terms of its heat kernel. The heat kernel admits at this stage an 
asymptotic expansion in powers of proper-time with coefficients which are 
known as Seeley-DeWitt coefficients \cite{DeWitt:1965,Seeley:1967}. In 
this way one can finally arrive at the expansion for the effective action 
at large distances. It is a very powerful method which is known as the  
background field method \cite{DeWitt:1975} and which allows the 
construction of the low-energy EFT action in the single-closed-loop 
approximation, when the underlying high energy theory (which in particular 
might be an EFT of some fundamental theory) is known. Many aspects of this 
approach in relation to chiral gauge theories are reviewed in 
\cite{Ball:1989}.

In a recent study \cite{Osipov:2001} of the low-energy structure of the NJL 
model with the linear realization of explicitly broken $SU(2)\times SU(2)$ 
chiral symmetry on the basis of the Schwinger-DeWitt proper-time formalism  
we came to the necessity of performing systematic resummations inside the 
proper-time expansion. These resummations occur when the one loop diagrams 
of the proper-time Green's function involve particles with different but 
still comparable masses. There we have obtained the first three terms
in the asymptotic expansion of the corresponding heat kernel, however at that 
time we could not suggest the general resummation procedure without which 
the method cannot be considered as complete. The aim of this letter is to 
present a general method for the construction of the heat kernel asymptotic 
expansion for such non-degenerate cases. The algorithm for resummations, 
Eq.(\ref{recfor}), is formulated on a purely algebraic basis and is 
completely novel. It leads us to an expansion which can be classified
as a series in inverse powers of mass with coefficient functions 
generalizing the standard Seeley-DeWitt coefficients. Let us clarify
this place. The incorporation of nontrivial mass matrices in the heat
kernel expansion is by no means a new, or unsolved problem. It has been
considered in detail, for example, in \cite{Lee:1989}. However, without the
resummation procedure the result cannot be cast in a chiral invariant form. 
We consider this symmetry property of asymptotic coefficients to be a 
crucial condition on any generalization of the Schwinger-DeWitt
result, which as it is well known fulfills this requirement. The resummations
come into play only after performing fully the integrations over the proper
time. It is important to note that in our approach at no instance do
we recur to the proper-time expansion. We do not expand in powers of
proper-time the mass dependent part of the heat kernel, for example,
by absorbing the mass term in the background fields. It is this
feature which makes our approach differ essentially from the ones (see 
for example \cite{Ven:1998,Gilkey:1998}) where authors study the proper-time 
asymptotics of heat kernels with arbitrary matrix-valued scalar
potentials, and thus nontrivial mass matrices in particular. 

Our starting point for calculations is the modulus of the functional 
fermion determinant for the one-loop effective action, given by the
proper-time integral
\begin{equation}
\label{logdet}  
  W[Y]=-\ln |\det D|=\frac{1}{2}\int^\infty_0\frac{dT}{T}\rho 
       (T,\Lambda^2)\mbox{Tr}\left(e^{-TD^\dagger D}\right),
\end{equation}
which can always be regularized by using, for example,
the Pauli-Villars cutoff \cite{Pauli:1949} incorporated through the 
kernel $\rho (T,\Lambda^2)$. We do not need the explicit form of $\rho 
(T,\Lambda^2)$ in the following. The calculation will be performed in 
Euclidean space. The elliptic operator $D^\dagger D$ has the form:
\begin{equation}
\label{DD}
      D^\dagger D=m^2+B, \quad B=-\partial^2 +Y,
\end{equation}
where $Y$ is a matrix-valued function of scalar and pseudoscalar background fields. In the most 
general case the mass term $m^2$ does not commute with $Y$. The first 
step is the evaluation of the heat kernel in a fictitious Hilbert space. 
We shall use here the formalism developed by Fujikawa \cite{Fujikawa:1979}.  
As a result we have
\begin{equation}
\label{logdet2}  
     W[Y]=\frac{1}{2}\int d^4x\int\frac{d^4p}{(2\pi )^4}
          \int^\infty_0\frac{dT}{T^3}\rho (T,\Lambda^2)
          e^{-p^2}\mbox{tr}\left(e^{-T(m^2+A)}\right)\cdot 1,
\end{equation}
where $A=B-2ip\partial /\sqrt{T}$ and tr is trace on the internal space. 
To simplify our consideration and make ideas more transparent let us 
choose the $SU(2)\times SU(2)$ group as a group of chiral transformations
acting on background fields. In this case the most general expression
for the square of the mass matrix is a sum  
\begin{equation}
\label{m2}
   m^2=K\tau_0 +M\tau_3, \quad K=\frac{1}{2}(m_u^2+m_d^2), \quad
                               M=\frac{1}{2}(m_u^2-m_d^2),
\end{equation}
with $\tau_0=1,\ \tau_i,\ (i=1,2,3)$ being the Pauli matrices for isospin.
Since $[m^2,Y]\ne 0$, we shall use the following operator identity, which is 
well-known in quantum mechanics, to factorize the mass matrix from the heat 
kernel in Eq.(\ref{logdet2}):    
\begin{equation}
\label{factor}
   \mbox{tr}\left(e^{-T(M\tau_3+A)}\right)=\mbox{tr}\left(
   e^{-TM\tau_3}\left[1+\sum^\infty_{n=1}(-1)^n f_n(T,A)\right]\right).
\end{equation}
Here $f_n(T,A)$ is equal to
\begin{equation}
\label{fnA}
   f_n(T,A)=\int^T_0ds_1\int^{s_1}_0ds_2\ldots
   \int^{s_{n-1}}_0ds_n A(s_1)A(s_2)\ldots A(s_n),
\end{equation}
where $A(s)=e^{sM\tau_3}Ae^{-sM\tau_3}$. If one takes into account
the permutation property of the trace operation in Eq.(\ref{factor}),
the expressions for $f_n(T,A)$ can be simplified. We find in this way   
\begin{equation}
\label{fn1}
   f_1(T,A)=TA,
\end{equation}
\begin{equation}
\label{fn2}
   f_2(T,A)=\frac{T^2}{4}\left[A^2+A\tau_3A\tau_3
           +c^{(2)}(T)(A^2-A\tau_3A\tau_3)\right],
\end{equation}
\begin{equation}
\label{fn3}
   f_3(T,A)=\frac{T^3}{8}\left[\frac{A}{3}\{A,\tau_3\}^2
           +c^{(3)}_1(T)A\{A,\tau_3\}[\tau_3,A]
           +c^{(3)}_2(T)(A^3-A\tau_3A\tau_3A)\right],
\end{equation}
\begin{eqnarray}
\label{fn4}
   f_4(T,A)&=&\frac{T^4}{128}\left(\frac{1}{3}\{A,\tau_3\}^4
            -c^{(4)}_1(T)[\tau_3,A]^2\{A,\tau_3\}^2
            -c^{(4)}_2(T)[\tau_3,A]\{A,\tau_3\}^2[\tau_3,A] 
             \right.\nonumber \\
           &+&\left. c^{(4)}_3(T)(2[\tau_3,A]\{A,\tau_3\}
                                   [\tau_3,A]\{A,\tau_3\}
                                  +[\tau_3,A]^4)\right).  
\end{eqnarray}
For the functions $c^{(i)}_j(T)$ we have  
\begin{equation}
\label{c2T}
   c^{(2)}(T)=\frac{1}{2T^2M^2}\left(e^{2TM\tau_3}-1-2TM\tau_3\right),
\end{equation}
\begin{equation}
\label{c31T}
   c^{(3)}_1(T)=\frac{\tau_3}{2T^3M^3}\left[1+TM\tau_3
               +(TM\tau_3-1)e^{2TM\tau_3}\right],
\end{equation}
\begin{equation}
\label{c32T}
   c^{(3)}_2(T)=\frac{\tau_3}{2T^3M^3}\left(e^{2TM\tau_3}    
               -1-2TM\tau_3-2T^2M^2\right),
\end{equation}
\begin{equation}
\label{c41T}
   c^{(4)}_1(T)=\frac{3}{2T^4M^4}\left(e^{2TM\tau_3}    
               -1-2TM\tau_3-2T^2M^2-\frac{4}{3}T^3M^3\tau_3\right),
\end{equation}
\begin{equation}
\label{c42T}
   c^{(4)}_2(T)=\frac{1}{2T^4M^4}\left[(3-4TM\tau_3+2T^2M^2)
                e^{2TM\tau_3}-3-2TM\tau_3\right],
\end{equation}
\begin{equation}
\label{c43T}
   c^{(4)}_3(T)=\frac{1}{2T^4M^4}\left[(2TM\tau_3-3)
                e^{2TM\tau_3}+3+4TM\tau_3+2T^2M^2\right].
\end{equation}
Now one can integrate over momentum in Eq.(\ref{logdet2}), which is a 
standard procedure \cite{Ball:1989}. Up to total derivatives, which 
can be omitted in the effective action, we obtain the expansion
\begin{equation}
\label{logdet3}  
   W[Y]=\int\frac{d^4x}{32\pi^2} 
        \int^\infty_0\frac{dT}{T^3}\rho (T,\Lambda^2)e^{-TK}
        \mbox{tr}\left(e^{-TM\tau_3}[1-f_1(T,B)+f_2(T,B)
        -\tilde{f}_3(T,B)+\ldots ]\right),
\end{equation}
where the last term, $\tilde{f}_3(T,B)$, contains the additional 
contributions of order $\sim T^3$, coming from $f_4(T,A)$:
\begin{equation}
   \tilde{f}_3(T,B)=f_3(T,B)+\frac{T^3}{48}\left(\{B,\tau_3\}
                   \{\partial^2B,\tau_3\}-3c^{(4)}_2(T)[\tau_3,B]
                   [\tau_3,\partial^2B]\right).
\end{equation} 

Let us replace in Eq.(\ref{logdet3}) the operator $B$ by its expression 
in terms of $Y$ (see Eq.(\ref{DD})). The result is
\begin{equation}
\label{logdet4}  
   W[Y]=\int\frac{d^4x}{32\pi^2}  
        \int^\infty_0\frac{dT}{T^3}\rho (T,\Lambda^2)e^{-TK}
        \mbox{tr}\left(e^{-TM\tau_3}[1-f_1(T,Y)+f_2(T,Y)
        -\bar{f}_3(T,Y)+\ldots ]\right),
\end{equation}
where $\bar{f}_3(T,Y)$ is equal to
\begin{equation}
   \bar{f}_3(T,Y)=f_3(T,Y)+\frac{T^3}{8}\left[\left(
                 \frac{1}{3}+2c^{(3)}_1-c^{(4)}_2\right)(\partial Y)^2 
                 +\left(\frac{1}{3}-2c^{(3)}_1+c^{(4)}_2\right)
                 (\tau_3\partial Y)^2\right].
\end{equation} 

The last step is to integrate over the proper-time $T$ in 
Eq.(\ref{logdet4}). The integrals over $T$ can be reduced to 
combinations of some set of elementary integrals $J_n(\mu^2)$
\begin{equation}
\label{Jn}
   J_n(\mu^2)=\int^\infty_0\frac{dT}{T^{2-n}}e^{-T\mu^2}\rho 
   (T,\Lambda^2),
\end{equation}
where $n$ is integer. Our task now is to find the algorithm which will 
automatically give a chiral invariant grouping for the background fields
as well as the mass dependent coefficients before them. To this end, 
it is necessary to reorganize the asymptotic series, given by 
Eq.(\ref{logdet4}), in the form
\begin{equation}
\label{logdet5} 
   W[Y]=\int\frac{d^4x}{32\pi^2}\sum^\infty_{i=0}I_{i-1}\mbox{tr}(a_i),
\end{equation}
where $2I_i\equiv J_i(K-M)+J_i(K+M)$. The necessary resummations inside 
the starting expansion (\ref{logdet4}) are determined by the recursion
relations
\begin{equation}
\label{recfor}
   J_i(K-M)-J_i(K+M)=\sum^\infty_{n=1}\frac{M^n}{n!}\left[
                     J_{i+n}(K+M)-(-1)^nJ_{i+n}(K-M)\right].  
\end{equation}
One can prove this useful identity for any integer $i$, calculating 
the first order derivative in $M$ in both sides of this equation. 
Passing from Eq.(\ref{logdet4}) to Eq.(\ref{logdet5}) corresponds to
choosing $J_i(K-M)+J_i(K+M)$ as the mass dependent factor in the 
asymptotic expansion, instead of $J_i(\mu^2)$ in the standard case of the
Schwinger-DeWitt series. The difference $J_i(K-M)-J_i(K+M)$ can be always
expressed through the infinite sum of $I_n$ with $n>i$, as it follows
from Eq.(\ref{recfor}). As a result of these manipulations one can 
find the coefficients $a_i$ in Eq.(\ref{logdet5}). The first five of them 
are equal to
$$
     a_0=1, \quad a_1=-Y, \quad a_2=\frac{Y^2}{2}+M\tau_3Y, \quad 
      a_3=-\frac{Y^3}{3!}-\frac{M}{2}\tau_3Y^2-\frac{1}{12}(\partial 
      Y)^2, 
$$
\begin{equation}
\label{coeff}
       a_4=\frac{Y^4}{4!}+\frac{M}{6}\tau_3Y^3-\frac{M^2}{12}(Y^2-Y\tau_3
       Y\tau_3)-\frac{M^3}{3}\tau_3Y+\frac{1}{12}(Y+M\tau_3)
       (\partial Y)^2+\frac{1}{120}(\partial^2 Y)^2.
\end{equation}

At this stage it can be verified that if the operator $D^\dagger D$ is 
defined to transform in the adjoint representation $\delta 
(D^\dagger D)=i[\omega,D^\dagger D]$, the coefficient functions $a_i$ are 
invariant under the global infinitesimal chiral transformations with 
parameters $\omega =\alpha +\gamma_5\beta$. This property extremely 
simplifies calculations, since it gives an alternative way to obtain $a_i$. 
Indeed, to find coefficient functions in Eq.(\ref{logdet5}) one can simply 
integrate the equation $\delta a_i=0$, using the corresponding Seeley-DeWitt 
coefficient as a starting point and constructing the necessary counterterms 
in such a way as to satisfy the aforementioned equation. 

The present result is in agreement with the standard Schwinger-DeWitt 
expansion, when $M=0$. For the case with $M\ne 0$ our formula 
(\ref{logdet5}) is a new asymptotic series which can be used to
construct the low-energy EFT action when the local vertices are 
induced by one-loop diagrams involving particles with different masses. 
It is a direct extension of the DeWitt WKB expansion constructed on the 
basis of the Schwinger proper-time method. Our approach can be used for 
a wide range of interesting applications, such as, for instance, the 
operator-product expansion \cite{Novikov:1984}, or derivative expansions 
\cite{Chan:1985}, or heat-kernel one-loop renormalizations 
\cite{Taron:1997}. It is not difficult to extend our method to the 
cases with more complicated symmetry groups. For instance, in the 
sequel to this letter we have already obtained asymptotic coefficients 
$a_i$ in the case of $SU(3)\times SU(3)$ chiral symmetry group 
\cite{Osipov:su3}. The other direction for development is to include 
minimal coupling of the loop particles to gauge fields.  

We would like to thank D. V. Vassilevich for clarifying correspondence
and valuable remarks. This work is supported by grants provided by 
Funda\c{c}\~{a}o para a Ci\^encia e a Tecnologia,
PRAXIS/C/FIS/12247/1998, PESO/P/PRO/15127/1999, POCTI/1999/FIS/35304, 
CERN/P/FIS/40119/2000 and NATO ``Outreach" Cooperation Program.

\baselineskip 12pt plus 2pt minus 2pt

\end{document}